\documentclass[aps,prd,12point,twocolumn,nofootinbib,showpacs,superscriptaddress]{revtex4-2}
\def\theequation{\arabic{section}.\arabic{equation}}
\usepackage{psfrag}
\usepackage{subfigure}
\usepackage{color}
\usepackage{mathrsfs}
\usepackage{graphicx}
\usepackage{amssymb, bm}
\usepackage{amsmath, amsthm}
\usepackage{epstopdf}
\usepackage{hyperref}
\usepackage{enumerate}
\usepackage{longtable}
\usepackage{amsmath}
\usepackage[normalem]{ulem}
\usepackage[title]{appendix}

\newcommand{\be}{\begin{equation}}
\newcommand{\ee}{\end{equation}}
\definecolor{pinegreen}{rgb}{0.0, 0.47, 0.44}

\begin{document}
\def\theequation{\arabic{section}.\arabic{equation}}

\title{Einstein gravity as the thermal equilibrium state of 
a nonminimally coupled scalar field geometry} 

\author{Numa Karolinski}
\email[]{nkarolinski22@ubishops.ca}
%\thanks{}
\affiliation{Department of Physics \& Astronomy, Bishop's University, 
2600 College Street, Sherbrooke, Qu\'ebec, Canada J1M~1Z7
}
\author{Valerio Faraoni}
\email[]{vfaraoni@ubishops.ca}
%\thanks{}
\affiliation{Department of Physics \& Astronomy, Bishop's University, 
2600 College Street, Sherbrooke, Qu\'ebec, Canada J1M~1Z7
}

%\date{\today}
\begin{abstract} 

We test ideas of the recently proposed first--order thermodynamics of 
scalar--tensor gravity using an exact geometry sourced by a conformally 
coupled scalar field. We report a non--monotonic behaviour of the 
effective ``temperature of gravity'' not observed before and due to a new 
term in the equation describing the relaxation of gravity toward its state 
of equilibrium, {\em i.e.}, Einstein gravity, showing a richer range of 
thermal behaviours of modified gravity than previously thought.

\end{abstract}

%\pacs{04.50.+h, 04.90.+e}
%alternative theories of gravity
%Other topics in GR and gravitation

%\keywords{thermodynamics of modified gravity, scalar-tensor gravity.}

\maketitle
% If in two-column mode, this environment will change to single-column
% format so that long equations can be displayed. Use
%\begin{widetext}
%\end{widetext}

\section{Introduction}
\label{sec:1}
\setcounter{equation}{0}

Einstein's theory of gravity, General Relativity (GR) has been very 
successful in the regimes in which it is tested 
\cite{Willbook,Will:2014kxa, Berti:2015itd,Baker:2014zba} but 
there is little doubt that, ultimately, 
it has to be replaced by some other theory (of which there is no  shortage \cite{Willbook, Clifton:2011jh, Heisenberg:2018vsk, 
Heisenberg:2018acv, CANTATA:2021ktz}). First, GR contains 
spacetime singularities inside black holes and in cosmology, thus 
predicting its own failure. These spacetime singularities should 
presumably be cured by quantum mechanics, but virtually all attempts to 
introduce quantum corrections also introduce deviations from GR 
\cite{Stelle:1976gc,Stelle:1977ry}. 
The low--energy limit of the simplest string theory, the bosonic string, 
does not reproduce GR but yields $\omega=-1$ Brans--Dicke gravity instead 
\cite{Callan:1985ia,Fradkin:1985ys}.

In addition to severe observational tensions 
\cite{Verde:2019ivm,DiValentino:2021izs}, from the theoretical point of 
view the standard GR--based $\Lambda$--Cold Dark Matter ($\Lambda$CDM) 
model of cosmology \cite{AmendolaTsujikawabook} is left wanting. Its main 
ingredient, {\em i.e.}, the dark energy accounting for approximately 70\% 
of the energy content of the universe, was introduced overnight to explain 
the current acceleration of the universe discovered in 1998 with Type Ia 
supernovae. The nature of this dark energy is a mystery. It is believed 
that, if one explains it with the cosmological constant $\Lambda$, extreme 
fine--tuning arises. An alternative to dark energy consists of modifying 
gravity at large scales \cite{Capozziello:2003tk,Carroll:2003wy}. For this 
purpose, $f(R)$ theories of gravity are very popular (see 
\cite{Sotiriou:2008rp,DeFelice:2010aj,Nojiri:2010wj} for reviews). The first scenario of inflation in 
the early universe, Starobinsky inflation \cite{Starobinsky:1980te}, which 
is also the scenario favoured by current observations \cite{Planck2}, is 
based on quadratic corrections to GR, where $f(R)=R+\alpha\, R^2$ (here $R$ denotes the Ricci scalar).

$f(R)$ gravity is a subclass of scalar--tensor gravity, the prototype of 
which is the original Brans--Dicke theory \cite{Brans:1961sx} later 
generalized by various authors \cite{Bergmann:1968ve, Nordtvedt:1968qs, 
Wagoner:1970vr, Nordtvedt:1970uv}. These ``old'' or ``first generation'' 
scalar--tensor theories were further generalized by Horndeski 
\cite{Horndeski}.  
The past decade has seen intense research activity on the 
rediscovered Horndeski theories, 
which were believed to be the most general scalar--tensor gravities with 
second order equations of motion, avoiding  the notorious 
Ostrogradsky instability that plagues theories with higher order 
equations. This record now belongs to further 
generalizations, the so--called Degenerate Higher Order Scalar--Tensor 
(DHOST) theories (see, {\em e.g.}, \cite{H1,H2,H3, GLPV1, GLPV2, 
DHOST1,DHOST2,DHOST3, DHOST4, DHOST5, DHOST6, DHOST7, DHOSTreview1, 
DHOSTreview2, Creminellietal18, Langloisetal18} 
and the references therein).

From another point of view, the idea has been proposed that perhaps, 
unlike 
the other three known interactions, gravity is not fundamental but could 
instead be emergent, similar to the way macroscopic thermodynamics emerges 
from microscopic degrees of freedom. This idea has been pursued in various 
implementations, see {\em e.g.}, \cite{Padmanabhan:2008wi, 
Padmanabhan:2009vy, Hu:2009jd, Verlinde:2010hp, Carlip:2012wa, 
Giusti:2019wdx}. One remarkable piece of work is Jacobson's thermodynamics 
of 
spacetime, in which the Einstein equation of GR is derived with purely 
thermal considerations and plays a role analogous to that of a macroscopic  
equation 
of state \cite{Jacobson:1995ab}. Furthermore, quadratic gravity (which is 
an $f(R)$, therefore a scalar--tensor, gravity) can be obtained in a 
similar way, but its derivation requires the introduction of  
entropy--generation terms \cite{Eling:2006aw}. The implication is that  GR 
is a thermal equilibrium state at zero temperature, while 
modified gravity corresponds to an excited state at higher temperature 
\cite{Eling:2006aw,Chirco:2010sw}.\footnote{This fact is natural when 
extra propagating degrees of freedom, in addition to the two massless 
spin--two modes 
of GR, are excited.} Problem is, the ``temperature of gravity'' and 
equations describing the relaxation to the GR equilibrium have never been 
found, in spite of a large literature on spacetime thermodynamics.

Recently, a more modest proposal was advanced in which this idea of 
modified gravity being an excitation of the GR equilibrium is re--examined 
in a context completely different from Jacobson's thermodynamics of 
spacetime. This new proposal, dubbed ``first--order thermodynamics of 
scalar--tensor gravity'' begins by noticing that the field equations of 
this class of theories can be written as effective Einstein equations with 
an effective stress--energy tensor in their right--hand 
side acting as an effective source, which has the structure of a 
dissipative fluid. This fact is well--known in special scalar--tensor 
theories or for special geometries (especially cosmological ones) 
\cite{Madsen:1988ph} 
and it extends to generic ``first--generation'' scalar--tensor theories 
\cite{Pimentel89,Faraoni:2018qdr} and to ``viable'' Horndeski 
\cite{Quiros:2019gai,Giusti:2021sku} gravity. Taking this 
dissipative structure seriously, one attempts to apply Eckart's 
first--order thermodynamics \cite{Eckart40} to it. Rather unexpectedly, 
the main constitutive relation of Eckart's theory (a generalized Fourier 
law) is satisfied \cite{Faraoni:2018qdr}, which makes it possible to read 
off this equation the product ${\cal KT}$ of a ``thermal conductivity of 
spacetime'' ${\cal K}$ and of the ``temperature of gravity'' ${\cal T}$ 
\cite{Faraoni:2018qdr, Faraoni:2021lfc, Faraoni:2021jri, Giusti:2021sku, 
Giusti:2022tgq,Faraoni:2023hwu,Miranda:2022wkz}. GR, obtained when the 
gravitational Brans--Dicke--like scalar field $\psi$ is constant, 
corresponds to ${\cal KT}=0$ while scalar--tensor gravity is a state at 
${\cal KT}>0$ \cite{Faraoni:2018qdr, Faraoni:2021lfc, Faraoni:2021jri, 
Giusti:2021sku}.

Thus far, this formalism is the closest that one has come to defining a 
``temperature of gravity''. An equation describing the approach to the GR 
thermal equilibrium, or the departure from it, is also provided in the 
formalism \cite{Faraoni:2021lfc, Faraoni:2021jri}, which is still  
under development. The formalism has been extended to ``viable'' Horndeski 
gravity \cite{Giusti:2021sku}, applied to cosmology 
\cite{Giardino:2022sdv,Faraoni:2023hwu}, to the Einstein frame description 
of these theories \cite{Faraoni:2022gry}, and to multi--scalar--tensor 
theories \cite{Miranda:2022uyk}. To make progress and gain insight into 
this new thermal description of gravity, one needs to test its basic ideas 
with special theories of modified gravity and their analytic solutions. 
While this 
work has begun \cite{Faraoni:2022doe,Faraoni:2022jyd,Faraoni:2022fxo, 
Giardino:2023qlu}, there are still many open questions and we continue 
this study here by applying the first--order thermodynamics to a special 
solution of nonminimally coupled scalar field theory (which is a 
scalar--tensor gravity) found recently by Sultana \cite{Sultana:2015lja}. 
This spacetime is inhomogeneous, spherically symmetric, and 
time--dependent 
and is conformal to a GR geometry obtained by Sultana by generalizing a 
previous GR solution due to Wyman \cite{Wyman81} to include a (positive) 
cosmological constant $\Lambda$ (we refer to the latter as the 
Sultana--Wyman solution).

In order to fix the notation, the next section recalls the basics of 
scalar--tensor gravity and of 
nonminimally coupled scalar field theory and introduces Sultana's 
solution, whose geometry has already been studied in detail in 
\cite{Faraoni:2022doe}. Section~\ref{sec:3} discusses the approach to the 
GR 
equilibrium identified by ${\cal KT}=0$. This discussion includes a term 
in the relevant relaxation equation that was set to zero for 
simplicity in previous literature. Section~\ref{sec:4} makes a brief 
parallel 
with an exact solution 
of Brans--Dicke theory conformal to the Sultana--Wyman solution, 
discussing its thermal properties.

We follow the notation of Ref.~\cite{Waldbook}: the signature of the 
metric tensor $g_{ab}$ is ${-}{+}{+}{+}$, units are used in which the 
speed of light $c$ and Newton's constant $G$ are unity (but, for 
convenience, we restore $G$ when discussing the nonminimally coupled 
scalar field), $\kappa \equiv 8\pi G$, $R_{ab}$ is the Ricci tensor, $R 
\equiv {R^a}_a$, $G_{ab} \equiv R_{ab}-R g_{ab}/2$ is the 
Einstein tensor, and $\Box \equiv g^{ab} \nabla_a\nabla_b $ is the curved 
space d'Alembertian.

\section{Scalar--tensor gravity and the Sultana 
solution}
\label{sec:2}
\setcounter{equation}{0}

The gravitational sector of ``first generation'' scalar--tensor gravity is 
described by the Jordan frame action \cite{Brans:1961sx, Bergmann:1968ve, 
Nordtvedt:1968qs,Wagoner:1970vr, Nordtvedt:1970uv} 
\begin{eqnarray}
S_\mathrm{ST} &=& \frac{1}{16\pi} \int d^4 x \, \sqrt{-g} \left[ \psi R 
-\frac{\omega(\psi) }{\psi} 
\, \nabla^c \psi \nabla_c \psi  -V(\psi) \right] \,, \nonumber\\
&&\label{STaction}
\end{eqnarray}
where $\psi$ is the Brans--Dicke--like scalar field corresponding, 
approximately, to the inverse of the effective gravitational coupling 
$G_\mathrm{eff}=1/\psi$,  $\omega (\psi)$ is the 
``Brans--Dicke coupling'' (which was constant in the original Brans--Dicke 
theory \cite{Brans:1961sx}), $V(\psi)$ is a scalar field potential, and 
$g$ is the determinant of the spacetime metric $g_{ab}$. The vacuum field 
equations 
obtained by varying the action~(\ref{STaction}) are  
\begin{eqnarray}
G_{ab} &=& \frac{\omega}{\psi^2}\Big(\nabla_a\psi\nabla_b\psi - 
\frac{1}{2}g_{ab}\nabla^c\psi\nabla_c\psi\Big) \nonumber\\
&&\nonumber\\
&\, & + \frac{1}{\psi}\Big(\nabla_a\nabla_b\psi - g_{ab}\Box\psi\Big) - 
\frac{V}{2\psi}g_{ab}  \,,
\end{eqnarray}

\be
\Box\psi = \frac{1}{2\omega+3} \left( \psi\, \frac{dV}{d\psi} -2V  
-\frac{d\omega}{d\psi} \nabla^c\psi \nabla_c\psi\right) \,. 
\label{Boxpsi}
\ee

By conformally rescaling the metric and redefining the scalar field 
according to
\begin{eqnarray}\label{met}
g_{ab}\rightarrow\tilde{g}_{ab}=\Omega^{2}g_{ab}=\psi \, g_{ab} \,,
\end{eqnarray}
\begin{eqnarray}\label{trans}
\psi \rightarrow \tilde{\psi}=
\sqrt{\frac{|2\omega+3|}{16\pi}} \ln\Big(\frac{\psi}{\psi_{\ast}}\Big)
\end{eqnarray}
(where $\psi_{\ast}$ is a positive constant), the Brans--Dicke
action is recast in its Einstein frame form (where quantities are denoted 
by a tilde)
\begin{eqnarray}
S_\mathrm{BD}&=&\int d^{4}x \sqrt{-\tilde{g}}  \,
\Big[\frac{\tilde{R}}{16\pi}-
\frac{1}{2} \, \tilde{g}^{ab}\,\tilde{\nabla}_{a}\tilde{\psi}\,
\tilde{\nabla}_{b}\tilde{\psi}-U(\tilde{\psi})\Big] \,,\nonumber\\
&&
\end{eqnarray}
where
\begin{eqnarray}\label{U}
U(\tilde{\psi})=\frac{V(\psi)}{\psi^{2}}\Big|_{\psi=\psi(\tilde{\psi})} 
\,.
\end{eqnarray}
In the Einstein conformal frame, vacuum scalar--tensor gravity looks like 
GR sourced by a minimally coupled scalar field $\tilde{\psi}$. The 
conformal transformation is commonly used as a  solution--generating 
technique using GR solutions sourced by minimally coupled scalar fields 
as seeds.

\subsection{Jordan frame scalar--tensor gravity}

The first step leading to the first--order thermodynamics of scalar--tensor gravity consists of writing the field equations as effective Einstein equations $G_{ab}= T_{ab}^{(\psi)}$ by collecting all terms other than the Einstein tensor $G_{ab}$ in the right--hand side. By assuming that $\nabla^a \psi$ is timelike and future--oriented, $T_{ab}^{(\psi)}$ necessarily has the structure of a dissipative fluid energy--momentum tensor \cite{Waldbook}
\be 
T_{ab}^{(\psi)}  =  \rho^{(\psi)}  u_a u_b + P^{(\psi)}  h_{ab} + 
\pi_{ab}^{(\psi)}  + q_a^{(\psi)}  u_b + q_b^{(\psi)}  u_a 
\,,\label{dissipativeTab}
\ee
with four--velocity 
\be
u^a=\frac{\nabla^a\psi}{\sqrt{ - \nabla^c\psi \nabla_c\psi}} \,,
\ee
spatial 3--metric
\be 
h_{ab}=u_a u_b + g_{ab} \,,
\label{3spacemetric}
\ee
energy density
\be
\rho^{(\psi)} = T_{ab}^{(\psi)} u^a u^b  \,, \label{mc_energydensity}
\ee
isotropic pressure
\be
P^{(\psi)} = \frac{1}{3} \, {{T^{(\psi)} }^a}_a \,, 
\label{mc_isotropicpressure}
\ee
heat flux density
\be
q_a^{(\psi)} = -T^{(\psi)}_{cd}u^c{h_a}^d  \,,  
\label{mc_heatfluxdensity}
\ee
and anisotropic stress tensor
\be
\pi_{ab}^{(\psi)} = T^{(\psi)}_{c d} h_a{ }^c h_b{ }^d - P h_{a b} 
\,. \label{mc_anisotropicstresstensor}
\ee
The isotropic pressure is decomposed into non--viscous and 
viscous contributions,
\be
P=\bar{P}+ P_\mathrm{visc} \,.
\ee

This dissipative fluid structure is common to all symmetric 2--index tensors and, {\em per se}, there is no physics in the decomposition~(\ref{dissipativeTab}) \cite{Faraoni:2023hwu}. However, by taking seriously the 
dissipative fluid structure of $T_{ab}^{(\psi)}$, one is tempted to apply Eckart's first--order thermodynamics 
\cite{Eckart40} to  it. This dissipative $\psi$--fluid does not, in general, satisfy energy conditions and it is impossible to identify unambigously all the thermodynamical quantities familiar from  real fluids. However, one can limit oneself to  considering the assumptions of Eckart's thermodynamics that relate heat flux density, anisotropic stresses, and viscous pressure to the kinematic quantities of the effective fluid, {\em i.e.}, the assumed constitutive relations where
\be
q_a =-{\cal K} h_{ab} \left( \nabla^b {\cal T}+ {\cal T} \dot{u}^b 
\right)  \,,\label{genFourier}
\ee
\be
\pi_{ab} = -2 \eta \, \sigma_{ab} \,, 
\ee
\be
P_\mathrm{visc} = -\zeta \, \Theta \,,
\ee
${\cal K}$ is the thermal conductivity, $\dot{u}^a \equiv u^c \nabla_c u^a$ is the fluid four--acceleration,  $\Theta \equiv \nabla_c u^c$ is the expansion scalar, and 
\be
\sigma_{ab}= {h_a}^c \,{h_b}^d \,\nabla_{(c} u_{d)} 
\label{shear_tensor_general_expression}
\ee 
is the shear tensor, while $\zeta$ and $\eta$ are effective bulk and shear viscosity coefficients. 
It is a miracle that the effective $\psi$--fluid satisfies Eckart's 
constitutive relation~(\ref{genFourier}) (with $h_{ab}\nabla^b {\cal 
T}=0$).  
This fact allows one to identify the product of a ``thermal conductivity'' 
${\cal K}$ and a ``temperature'' ${\cal T}$ of gravity as 
\cite{Faraoni:2021lfc,Faraoni:2021jri}
\be
{\cal KT}= \frac{ \sqrt{ -\nabla^c\psi \nabla_c\psi}}{\kappa \psi} 
\,.\label{generalKT}
\ee
GR, obtained for $\psi=$~const., corresponds to thermal equilibrium at 
${\cal KT}=0$. The approach to (or departure from) this GR 
equilibrium is described by \cite{Faraoni:2021jri}
\be
\frac{d \left( {\cal KT}\right)}{d \bar{\tau}} = \kappa \left( {\cal 
KT}\right)^2 -\Theta {\cal  KT} +\frac{\Box \psi}{\kappa \psi} 
\,,\label{KTevolution}
\ee
where $ \bar{\tau}$ is the proper time along the lines of the 
$\psi$--fluid and $d/d\bar{\tau} \equiv u^c \nabla_c  $.  The 
expansion scalar is \cite{Faraoni:2018qdr}
\begin{eqnarray}
    \Theta &=& \nabla _a u^a = \frac{1}{\sqrt{-\nabla_c \psi \nabla ^c \psi}} \left( \Box \psi - \frac{\nabla ^a \psi \nabla ^b \psi \nabla _a \nabla_b \psi}{\nabla _e \psi \nabla ^e \psi} \right) \,.\nonumber\\
    &&\label{expansion}
\end{eqnarray}

\subsection{Nonminimally coupled scalar fields}

Let us come to nonminimally coupled scalar fields. The gravitational 
sector of nonminimally coupled scalar field theory is described by the 
Jordan frame action 
\begin{eqnarray}
S_\mathrm{\small{NMC}} = \int d^4 x \, \sqrt{-g} &\bigg[& 
\bigg( \frac{1}{\kappa}- \xi \phi^2  \bigg)\frac{R}{2} \nonumber\\
&\,&\nonumber\\
&-& \frac{1}{2} \, \nabla^c\phi \nabla_c\phi -V(\phi) \bigg] \,,\label{SNMC}
\end{eqnarray} 
where the constant $\xi$ describes the nonminimal coupling of the scalar 
$\phi$ to the Ricci scalar. The corresponding field equations read 
\begin{eqnarray}
G_{ab} &=& \kappa \big( 1-\kappa\xi\phi^2\big)^{-1}  
\bigg( \nabla_a\phi\nabla_b\phi - \frac{1}{2}\, g_{ab}\nabla^c\phi 
\nabla_c\phi \nonumber\\ 
&&\nonumber\\
&\, &  - V g_{ab}+\xi \Big[ g_{ab}\,\square \big(\phi^2\big) 
-\nabla_a\nabla_b \big(\phi^2\big) \Big] \bigg) \,,
\end{eqnarray}
\be 
\Box \phi -\xi R\phi -\frac{dV}{d\phi} =0 
\,.\label{NMCeq} 
\ee 
Nonminimal coupling ({\em i.e.}, $\xi\neq 0$) seems 
to have been originally introduced in the context of radiation problems in 
curved space \cite{ChernikovTagirov}. It is unavoidable when quantizing a 
scalar field on a curved background (\cite{Callan:1970ze}, see also 
\cite{Birrell:1982ix,Birrell:1979ip,Nelson:1982kt, 
Ford:1981xj,Parker:1983pe}). The value $\xi=1/6$ (conformal 
coupling) 
makes Eq.~(\ref{NMCeq}) conformally invariant if $V(\phi)$ is  
quartic or identically zero.  From a classical point of view, 
conformal 
coupling is necessary to avoid causal pathologies, {\em i.e.}, the 
propagation of a massive scalar field $\phi$ strictly along the light cone 
\cite{Sonego:1993fw}. $\xi=1/6$ is also an infrared fixed point of the 
renormalization group in Grand Unified Theories 
\cite{Buchbinder:1985ba,Buchbinder:1985ew, Markov86, 
Odintsov:1990mt,Muta:1991mw, Elizalde:1994im, Buchbinderbook}.

The field equations for the conformally coupled $\phi$ can be written as 
\be
G_{ab} = G_\mathrm{eff} T_{ab}^{(\psi)} \,,
\ee
where 
\be
G_\mathrm{eff} = \frac{G}{1-\alpha^2 \phi^2} \,, \quad 
\alpha \equiv \sqrt{\frac{\kappa}{6}}  \label{Geff} 
\ee
is the effective gravitational coupling strength.  The effective fluid 
quantities derived from $T_{ab}^{(\phi)}$ are 
\begin{eqnarray}
\rho^{(\psi)} &=& T_{ab}^{(\psi)} u^a u^b = 
\bigg( 1- \frac{4\,\pi\,\phi^2}{3} \bigg)^{-1} \Bigg\{ -\frac{1}{2} \, 
\nabla^e 
\phi \nabla_e \phi  \nonumber\\ 
&&\nonumber\\
&\, &  +\frac{V(\phi)}{2} +\frac{1}{6}\bigg[ \frac{\nabla^a \phi 
\nabla^b \phi \nabla_a \nabla_b\left(\phi^2\right)}{\nabla^e \phi 
\nabla_e \phi} -\square\left(\phi^2\right)\bigg]\Bigg\}  \,, \nonumber\\
&\, &\label{energydensity_definition}
\end{eqnarray}
\begin{eqnarray}
P^{(\psi)} &=& \frac{1}{3} \, {{T^{(\psi)} }^a}_a  =  
\bigg(1-\frac{4\,\pi\,\phi^2}{3}\bigg)^{-1}\Bigg\{ -\frac{1}{2} \nabla^e 
\phi \nabla_e \phi  \nonumber\\ 
&&\nonumber\\
&\, & -\frac{V(\phi)}{2} +\frac{1}{18}\bigg[\frac{\nabla^a \phi 
\nabla^b \phi \nabla_a \nabla_b\left(\phi^2\right)}{\nabla^e \phi \nabla_e 
\phi} + 2\square\left(\phi^2\right)\bigg]\Bigg\} \,, \nonumber\\
&\, &\label{isotropicpressure_definition}
\end{eqnarray}
\begin{eqnarray}
q_a^{(\psi)} &=&  -T^{(\psi)}_{cd}u^c{h_a}^d = 
\frac{\Big(1-4\,\pi\,\phi^2/3 \Big)^{-1} 
\nabla^c\phi\nabla^d\phi}{6\,\big(-\nabla^e\phi\nabla_e \phi\big)^{3/2}} 
\,  \nonumber\\ 
&& \nonumber\\
&\, & \Big[ \nabla_d\phi\nabla_a\nabla_c\big(\phi^2\big) 
-\nabla_a\phi\nabla_c\nabla_d\big(\phi^2\big) \Big] \,,  
\label{heatflux_definition}
\end{eqnarray}
and
\begin{eqnarray}
\pi_{ab}^{(\psi)} &=& T^{(\psi)}_{c d} h_a{ }^c h_b{ }^d -  P h_{a 
b} =  -\frac{\Big(1- 4\,\pi\,\phi^2/3 \Big)^{-1}}{6\,\nabla^e 
\phi\nabla_e\phi}   \nonumber\\ 
&& \nonumber\\
&\, & \times \Bigg\{ \frac{1}{3} \, \big(\nabla_a\phi\nabla_b\phi  - 
g_{ab}\nabla^e\phi\nabla_e\phi\big)\bigg[\square\big(\phi^2\big) 
\nonumber\\
&& \nonumber\\
&\, &  - \frac{\nabla^c\phi\nabla^d\phi\nabla_c\nabla_d 
\big(\phi^2\big)}{\nabla^e\phi\nabla_e\phi}\bigg] + \nabla^d\phi 
\bigg[\nabla_d\phi\nabla_a\nabla_b\big(\phi^2\big) \nonumber\\
&& \nonumber\\
&\, & - \nabla_b\phi\nabla_a\nabla_d\big(\phi^2\big) 
- \nabla_a\phi\nabla_b\nabla_d\big(\phi^2\big) \nonumber\\
&&\nonumber\\
&\, & + 
\frac{\nabla^c\phi\nabla_a\phi\nabla_b\phi\nabla_c\nabla_d\big 
(\phi^2\big)}{\nabla^e\phi\nabla_e\phi} \bigg] \Bigg\} \,. \nonumber \\
&\, &\label{anisotropicstress_definition}
\end{eqnarray}

Here we are interested in a particular solution for a conformally coupled 
scalar field to elucidate features of the first--order thermodynamics of 
scalar--tensor gravity. The starting point is a solution of GR with a 
minimally coupled ({\em i.e.}, $\xi=0$) scalar field found by 
Wyman\footnote{This is sometimes 
called Wyman's ``other'' solution to distinguish it from the better--known 
solution discovered by Fisher \cite{Fisher:1948yn} and rediscovered over 
and over again \cite{BL57,JNW68, Buchdahl72, Wyman81}. The latter is the 
general solution of the $\Lambda=0$ Einstein equations which is static, 
spherically symmetric, asymptotically flat, and with a free scalar field 
as the source (see the recent review \cite{Faraoni:2021nhi}).} 
\cite{Wyman81}
\be
d\tilde{s}^2 = -\kappa r^2 dt^2 +2dr^2 + r^2 d\Omega_{(2)}^2 \,,
\ee
\be
\tilde{\phi}(t) = \tilde{\phi}_0 \, t \,,
\ee
where $d\Omega_{(2)}^2 \equiv d\vartheta^2 + \sin^2 \vartheta \, 
d\varphi^2 $ is the line element on the unit 2--sphere and 
$\tilde{\phi}_0$ is a constant. 
Wyman's ``other'' solution coincides with a special case of interior 
solutions for relativistic stars with a stiff fluid found by Iba\~nez \& 
Sanz \cite{IbanezSanz} in 1982, previously studied by Buchdahl \& Land 
\cite{BuchdahlLand}. This is, in turn, a special case of the Tolman~IV 
class 
of solutions of the Einstein equations discovered in 1939 \cite{Tolman39, 
Delgaty:1998uy, Stephanietal}. The matching with an exterior GR solution 
was studied in \cite{Faraoni:2021vpn}.

The Sultana geometry with $\Lambda>0$  \cite{Sultana:2015lja} 
coincides with another special case of a class of perfect fluid solutions 
given by Iba\~nez \& Sanz. It should probably be called 
Buchdahl-Land-Sultana-Wyman-Iba\~nez-Sanz (BLSWIS) solution. More 
precisely, this BLSWIS metric is a special limit of Buchdahl \& 
Land's \cite{BuchdahlLand} stiff fluid solution of the Einstein 
equations with vanishing cosmological constant but pressure 
\be
P= \rho-\rho_0 
\ee
(where $\rho$ is the fluid energy density and $\rho_0$ is a constant), 
which is supposed to describe an incompressible fluid but, in practice, 
a 
cosmological constant is re--introduced.
 
The more general Buchdahl--Land solution constitutes a special case of the 
Tolman~IV class of solutions \cite{Tolman39} describing the interior of a 
perfect fluid ball with cosmological constant \cite{Delgaty:1998uy}.

Sultana \cite{Sultana:2015lja} generalized the Wyman solution to the case 
in which there is a positive cosmological constant $\Lambda$. This 
geometry (that we refer to as the Sultana--Wyman solution of GR) can be 
regarded as the Einstein frame version of a Jordan frame solution of 
conformally coupled scalar field theory. Then, the inverse conformal map 
from the Einstein to the Jordan frame produces a new vacuum solution of 
this theory (which is, of course, conformal to the Sultana--Wyman solution 
of GR), here referred to as the Sultana solution.

The inhomogeneous, spherically symmetric, and time--dependent Sultana 
solution of conformally coupled scalar field theory is 
\cite{Sultana:2015lja}
\begin{eqnarray}
    ds^2 &=& \cosh^2 (\alpha t) \left[ -\kappa r^2 dt^2 +\frac{2dr^2}{1-2\Lambda r^2/3} + r^2 d\Omega_{(2)}^2 \right] \,,\nonumber\\
    && \label{Sultanametric}
\end{eqnarray}
\be
\phi(t) = \pm \frac{1}{\alpha} \, \tanh (\alpha t) \,,\label{Sultanaphi}
\ee
where  $-\infty< t < +\infty$ and $ 0 < r< \sqrt{ \frac{3}{2\Lambda}}$.   
The scalar field potential obtained by mapping back the 
cosmological constant $\Lambda$ from the Einstein to the Jordan conformal 
frame is the Higgs potential \cite{Sultana:2015lja}
\be
V(\phi) =\frac{\Lambda}{\kappa} \left( 1 -\alpha^2 \phi^2 \right)^2 
 \,.\label{Higgs}
\ee
In general, the conformal map between Jordan and Einstein frames produces 
scalar field potentials that are not physically motivated, but this is not 
the case here. $V(\phi) $ is non--negative since (following 
\cite{Sultana:2015lja}) we assume $\Lambda \geq 0$. The effective 
gravitational coupling strength~(\ref{Geff}) is kept positive by requiring 
that $-\phi_c < \phi < \phi_c$, where $\phi_c \equiv \alpha^{-1}= \sqrt{ 
6/\kappa}$ is a critical scalar field value. With this potential, the 
action~(\ref{SNMC}) is invariant under the exchange $\phi \to -\phi$.

The Sultana geometry exhibits a spacetime singularity ar $r=0$, where the 
Kretschmann scalar
\be
R_{abcd} R^{abcd} = \frac{ 3+ 2 \cosh(\alpha t) + \cosh(4\alpha t)}{3r^4 
\cosh^8(\alpha t) }
\ee
diverges \cite{Sultana:2015lja}. 

The nature of the Sultana--Wyman solution of GR was studied in detail in 
Ref.~\cite{Banijamali:2019gry}, which analyzes its radial null geodesics. 
Since the Sultana geometry~(\ref{Sultanametric}), (\ref{Sultanaphi}) is 
conformal to the Sultana--Wyman geometry of GR its causal structure, which 
is conformally invariant, is the same. In particular, the singularity at 
$r=0$, where the curvature invariants diverge, persists. This is a naked 
singularity because no apparent horizons cover it. The equation locating 
all the apparent horizons, when they exist, is $\nabla^c {\cal R} \nabla_c 
{\cal R}=0$, where ${\cal R}$ is the areal radius \cite{Faraoni:2015ula} 
and this equation does not admit solutions in the Sultana spacetime  
\cite{Banijamali:2019gry}. 

The gradient of the  scalar field~(\ref{Sultanaphi})  
\be
\nabla^a \phi = \mp \frac{ {\delta_0}^a }{\kappa r^2 \cosh^4(\alpha t)} 
\ee 
is timelike, 
\be
\nabla^a \phi\nabla_a\phi = -\frac{1}{\kappa r^2 \cosh^6( \alpha t)} <0 
\ee
for any time $t$, but is not future--oriented (and, therefore, cannot be 
used to define an effective fluid four--velocity) unless the negative sign 
is adopted in Eq.~(\ref{Sultanaphi}), which we do here:\footnote{A 
possible alternative consists of defining $u^a \equiv - \left( 
-\nabla^c\phi \nabla_c \phi \right)^{-1/2} \nabla^a \phi$ 
\cite{Giusti:2022tgq}.} hence from now on $\phi(t) = -\alpha^{-1} 
\tanh(\alpha t)$ and $\psi=1-\alpha^2\phi^2 $. 
With this choice, the four--velocity of the effective 
$\phi$--fluid
\be
u^a = \frac{\nabla^a\psi}{ \sqrt{-\nabla^e\psi\nabla_e \psi}}
=  \frac{ {\delta_0}^a}{\sqrt{\kappa} \, r \cosh(\alpha t)}
\ee
is timelike, future--oriented, and normalized to $u^c u_c=-1$ at all times $t$ and coincides with the time direction of the observers comoving with the effective fluid. The three--dimensional metric on the 3--space orthogonal to $u^a$ is $h_{ab}$, described in Eq.~(\ref{3spacemetric}), where $ h_{ab}u^a =h_{ab} u^b=0$. For convenience, we report below the kinematic quantities defined by $u^a$. 

By using the Christoffel symbols 
\begin{eqnarray}
\Gamma^0_{00} &=& \alpha \, \tanh(\alpha t) \,,\\
&&\nonumber\\
\Gamma^1_{00} &=& \frac{\kappa \, r}{2} \left( 1-\frac{2\Lambda r^2}{3} 
\right)  \,,\\
&&\nonumber\\ 
\Gamma^2_{00} &=& \Gamma^3_{00} = 0\,,
\end{eqnarray}
the effective fluid four--acceleration 
\begin{eqnarray}
    \dot{u}^a &=& u^c \nabla_c u^a = u^0 \left( \partial_0 u^a + \Gamma^a_{00} u^0 \right)\nonumber\\
    &=& {\delta_1}^a \, \frac{ \left( 1-2\Lambda r^2/3\right) }{2r\cosh^2 ( \alpha t) }\,,
\end{eqnarray}
turns out to be purely radial as expected.  The velocity gradient (twice) projected onto the three--space orthogonal to $u^a$  decomposes as 
\be
V_{ab} \equiv {h_a}^c \, {h_b}^d \nabla_d u_c = \sigma_{ab} + 
\frac{\Theta}{3} \, h_{ab}\,,
\ee
where $\Theta$ is the expansion scalar, $\sigma_{ab}$ is the trace--free 
shear tensor, and the antisymmetric part of $V_{ab}$, the vorticity 
($\omega_{ab}$), vanishes because $u^a$ is derived from a scalar and is 
orthogonal to the hypersurfaces of constant $\phi$ and constant time $t$. 
Four--acceleration and shear are purely spatial,
$\dot{u}_c u^c = \sigma_{ab}u^a =\sigma_{ab} u^b =0$  
and ${\pi^a}_a=0$. 

The shear tensor and expansion scalar in 
coordinates $\left( t, r, \vartheta, \varphi \right)$ are (see 
Appendix~\ref{Appendix:A}) 
\be
\sigma_{ab} = 0 \,,
\ee
\be 
\Theta= \frac{3\, \tanh(\alpha t)}{\sqrt{6}\, r\, \cosh(\alpha 
t)} \,.
\ee

Next, one computes the effective fluid quantities composing the 
stress--energy tensor~(\ref{dissipativeTab}), obtaining (cf.  
Appendix~\ref{Appendix:A})
\begin{eqnarray}
\rho^{(\psi)}& =& \frac{\cosh(2\alpha t)\, \text{sech}^4(\alpha 
t)}{2\, \kappa\, r^2}  + \frac{\Lambda\, \text{sech}^2(\alpha t)}{2\, 
\kappa} \,,\\
&&\nonumber\\
P^{(\psi)}& =& \frac{\cosh(2\alpha t)\, \text{sech}^4(\alpha t)}{6\, 
\kappa\, r^2} -  \frac{\Lambda\, \text{sech}^2(\alpha t)}{2\, 
\kappa} \,,\\
&&\nonumber\\
q_a^{(\psi)}& =& {\delta_a}^1\, \frac{2\,\tanh (\alpha t)}{ 
\sqrt{6}\,\kappa\,r^2\,\cosh(\alpha t)} \,,\\
&&\nonumber\\
\pi_{ab}^{(\psi)}& =& 0 \,.
\end{eqnarray}
Furthermore, the viscous pressure reads 
\be
P^{(\psi)}_\mathrm{visc}=  \frac{\sinh^2(\alpha t) - 
1}{3\,\kappa\,r^2\,\cosh^4(\alpha t)} \,.
\ee

\section{First--order thermodynamics and approach to the GR 
equilibrium}
\label{sec:3}
\setcounter{equation}{0}

As for any first--generation scalar--tensor theory, Eckart's generalization~(\ref{genFourier}) of the Fourier law is satisfied, giving
\begin{eqnarray}
    q^a_{(\psi)} &=& -{\cal K}h_{ab} \left( \nabla^b {\cal T}+{\cal T}\dot{u}^b \right) = \frac{2 \sinh( \alpha t)}{\sqrt{6} \, \kappa r \cosh^2 (\alpha t)} \,  \dot{u}^a \,, \nonumber\\
    &&
\end{eqnarray}
which yields 
\be
{\cal KT} = - \frac{2\sinh (\alpha t)}{\sqrt{6} \, \kappa r \cosh^2 
(\alpha t)} \,.
\ee
${\cal KT}$ is positive--definite for $t<0$, therefore we restrict our 
consideration to negative times in the following. It is convenient to 
introduce a new time coordinate $\tau$ defined by $ d\tau \equiv  \cosh( 
\alpha 
t ) dt$, or
\be
\tau(t) = \frac{\sinh( \alpha t)}{\alpha} \,,
\ee
which is always well--defined since $d\tau/dt>0 $ at all times. In terms of $\tau$, the Sultana solution reads
\begin{eqnarray}
    ds^2 &=& -\kappa r^2 d\tau^2 + \left( 1+\alpha^2 \tau^2 \right) \left(\frac{ 2dr^2}{1-2\Lambda r^2/3} + r^2 d\Omega_{(2)}^2 \right) \,,\nonumber\\
    &&\label{Sultanametric2}
\end{eqnarray}
\be
\phi( \tau ) = - \frac{\tau }{ \sqrt{ 1+\alpha^2 \tau^2}} \,, \label{Sultanaphi2}
\ee
while
\be
{\cal KT} = \frac{-2\alpha \, \tau}{\sqrt{6} \, \kappa \, r \left( 
1+\alpha^2 \tau^2 \right)} \,. \label{KTintau}
\ee
\begin{figure}
    \centering
    \includegraphics[width=0.4833\textwidth]{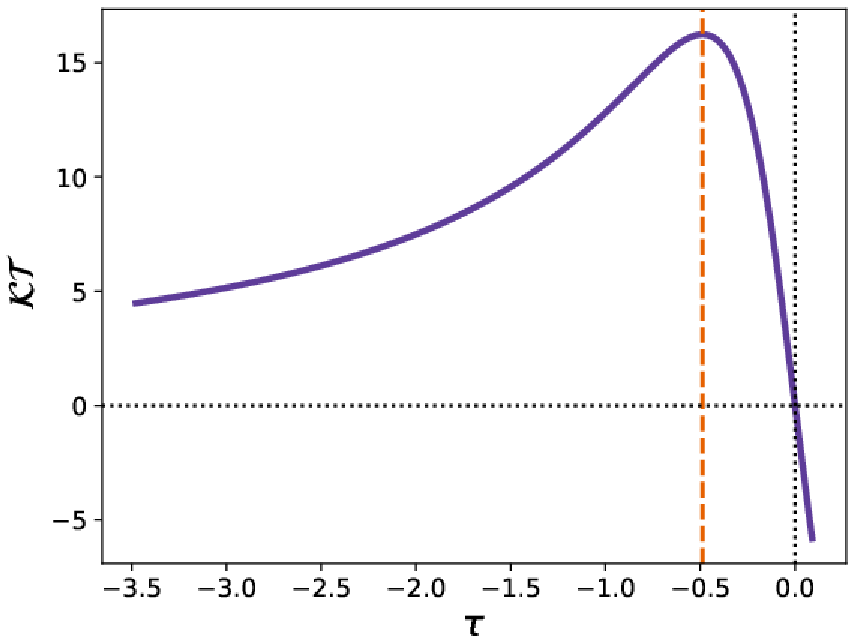}
    \caption{Evolution of $\mathcal{KT}$ with the time 
$\tau$ (in units in which $r=1$ and $G=1)$. The solid (purple) curve represents $\mathcal{KT}(\tau, 1)$. The dashed vertical (orange) line marks the maximum value of $\mathcal{KT}(\tau, 1)$.}
    \label{KT_scaling}
\end{figure}
Using $\psi=1-\alpha^2 \phi^2 =\left(1+\alpha^2\tau^2\right)^{-1}$, it is 
straightforward to see that Eq.~(\ref{KTintau}) matches the general expression~(\ref{generalKT}) of ${\cal KT}$ in first--generation scalar--tensor gravity. Furthermore, one has 
\begin{eqnarray}
    \Box \phi=0
\end{eqnarray}
and 
\begin{eqnarray}
R &=& \frac{- 4 \kappa V}{1-\alpha^2 \phi^2} = - 4 \Lambda  \left(1-\alpha^2 \phi^2 \right) = - \frac{4 \Lambda}{ \cosh^2 \left( \alpha \, t \right)} \,,\nonumber\\
&&
\end{eqnarray}
consistent with the equation of motion of $\phi$. However, $\Box \psi \neq 0$ and the last term on the right--hand side of Eq.~(\ref{KTevolution}) does not vanish, allowing for a more generic test of the basic ideas of Refs.~\cite{Faraoni:2018qdr, Faraoni:2021lfc, Faraoni:2021jri, Giusti:2021sku, Giusti:2022tgq, Faraoni:2023hwu, Miranda:2022wkz}.

As is clear from Eq.~(\ref{KTintau}), ${\cal KT} \to +\infty$ as $r\to 0^{+}$: the naked spacetime singularity at $r=0$ is ``hot'', in the sense that gravity departs from GR there, and the deviation is infinite. In the infinite past, $\tau\to-\infty$ (also $t\to -\infty$), $\phi(\tau)$ approaches a constant and ${\cal KT}$ tends to zero, thus gravity approaches GR asymptotically (Fig.~\ref{KT_scaling}). This fact is consistent with the idea that gravity ``cools'' as 3--space expands. In fact, the finite volume of 3--space is
\begin{eqnarray}
V^{(3)}(\tau&&) = \int d^3 x \, \sqrt{ g^{(3)} (\tau)} \nonumber\\
&&\nonumber\\
&&= \int_0^{\sqrt{\frac{3}{2\Lambda}}} dr\int_0^{\pi} d\vartheta 
\int_0^{2\pi} 
d\varphi  \sqrt{  \frac{ 2 \left(1+\alpha^2 \tau^2\right)^3 }{ 1-2\Lambda 
r^2 
/3}} \, r^2\sin\vartheta \nonumber\\
&&\nonumber\\
&&= 4\pi \sqrt{2}  \left( 1+\alpha^2 \tau^2 \right)^{3/2} 
\int_0^{\sqrt{\frac{3}{2\Lambda}}} dr \, \frac{r^2}{ \sqrt{ 1-2\Lambda 
r^2/3}} \nonumber\\
&&\nonumber\\
&&=  
\frac{3\sqrt{3} \pi^2}{2\Lambda^{3/2} } \left(1+\alpha^2 \tau^2 
\right)^{3/2} \to +\infty \quad \quad \mbox{as }\; \tau\to -\infty 
\,.\nonumber\\
&& 
\end{eqnarray} 
The 3--volume $V^{(3)}$ is infinite in the infinite past $\tau\to 
-\infty$, decreases 
monotonically for $\tau<0$, and reaches its absolute minimum 
$V_\mathrm{min} = \frac{3\sqrt{3} \pi^2}{2\Lambda^{3/2} } $ at $\tau=0$, 
then increases monotonically, diverging again as $\tau \to 
+\infty$. 

Since ${\cal KT}\to 0^{+}$ and $V^{(3)} \to +\infty$ as $\tau\to-\infty$, 
the Sultana solution corroborates the idea that the expansion of space 
``cools'' gravity, even when the third term $\frac{ \Box \psi}{\kappa\, 
\psi}$ in Eq.~(\ref{KTevolution}) does not vanish and for 
non--cosmological 
spacetimes in which 3--space still expands. The effective gravitational 
coupling $ G_\mathrm{eff}=\frac{G}{1-\left( \phi/\phi_c \right)^2} \to 
+\infty$ as $\tau\to-\infty$ and $\phi\to -\phi_c$. Thus far, based on 
previous exact solution examples, singularities of $G_\mathrm{eff}$ have 
been regarded on par with spacetime singularities where gravity becomes 
``hot''. As the Sultana solution shows, this picture is at least 
incomplete because here $G_\mathrm{eff}$, which depends only on time, 
diverges where 3--space expands without limit, but gravity ``cools'' 
instead. The effect of 3--space expansion dominates over divergences of 
the gravitational coupling. In the end, it is the product $G_\mathrm{eff} 
T_{ab}^{(\phi)}$ that enters the right--hand side of the field equations, 
so whether gravity approaches GR or departs from it is determined by the 
vanishing of this product and not by the individual factors 
$G_\mathrm{eff}$ and $T_{ab}^{(\phi)}$.

Increasing ${\cal KT}$ means increasing deviation from GR: we have 
\be
\frac{d \left( {\cal KT} \right)}{d\bar{\tau}} \equiv u^c \nabla_c 
\left( {\cal  KT} \right) = u^{\tau} \, \frac{ d\left( {\cal KT} 
\right)}{d\tau} \,,
\ee
where $\bar{\tau}$ is the proper time along the fluid lines of the 
effective $\phi$--fluid. The normalization $u^c u_c=-1$ (or, 
alternatively, the transformation property $u^{\tau}=\frac{\partial 
\tau}{\partial t} \, u^t$) gives  $u^{\tau}= 1/\left( \sqrt{\kappa} \, r 
\right)$ and 
\be
\frac{d \left( {\cal KT}\right)}{d\bar{\tau}} = 
\frac{1}{\sqrt{\kappa} \, r} \, \frac{d\left( {\cal KT}\right)}{d\tau}= 
\frac{\alpha^2 \tau^2 -1}{ 
3 \sqrt{\kappa}\,  r \left( 1+\alpha^2 \tau^2 \right)^2} \,.
\ee
${\cal KT}(\tau, r)$ increases for $-\infty < \tau < -\alpha^{-1}$, 
where it has a maximum ${\cal KT}_\mathrm{max} = \frac{1}{\sqrt{6} \, 
\kappa \, r} $ and decreases for $-\alpha^{-1}< \tau <0$, vanishing as 
$\tau\to 0^{-}$ and changing concavity at $\tau =- \sqrt{3}\alpha$. 
Therefore, gravity is extremely close to GR in the far past, then it 
gradually departs from it but only to a finite extent (${\cal KT}$ 
remains finite at all radii $r>0$), then approaches GR again, coinciding 
with it at $\tau=0$. To the best of our knowledge, all analytic solutions 
of scalar--tensor gravity studied thus far exhibit instead a monotonic 
approach to, or departure from, the GR equilibrium state.

The third term $ 
\frac{\Box\psi}{\kappa \psi} $ in the right--hand side of  
Eq.~(\ref{KTevolution}) is responsible for the non--monotonic 
behaviour of ${\cal KT}$ and dominates near $\tau=0$. It is possible for 
gravity to depart from GR and 
return to it after a temporary deviation, a behaviour that was not 
observed  before in the literature, which was limited to examples 
in which 
$\Box\psi=0$. 

\section{A related solution of Brans--Dicke and $f(R)$ gravity}
\label{sec:4}
\setcounter{equation}{0}

It is useful to contrast the thermal behaviour of gravity in the Sultana 
geometry with the one of other solutions in which $\Box\psi=0$. To this 
end, we choose a recent solution of Brans--Dicke theory closely related to 
the Sultana spacetime of the previous sections, for which the thermal 
evolution of gravity has not been discussed.

A family of solutions of Brans--Dicke gravity was generated in 
Ref.~\cite{Banijamali:2019gry} using the conformal transformation from 
Einstein to Jordan frame and the Sultana--Wyman geometry as a seed, 
obtaining  
\begin{eqnarray}
ds^{2} &=& -\kappa
r^{2}d\tau^{2}+\bigg(1-\frac{\tau}{\tau_{\ast}}\bigg)^{2}
\Bigg(\frac{2dr^{2}}{1- 2\Lambda r^2/3 }+r^{2}d\Omega^{2}_{(2)}\Bigg) \,, 
\nonumber\\
&& \label{new1}\\
\psi(\tau) &=&  \frac{\psi_{\ast}}{\Big(1- \tau/\tau_{\ast} \Big)^2 }  
\,, \label{new2}
\end{eqnarray}
where $\tau_{\ast}$ and $ \psi_{\ast}$ are constants related to the initial 
conditions. The Jordan frame scalar field  potential is the simple mass 
term
\be
V(\psi) = \frac{m^2\psi^2}{2} \,, \quad \quad m^2=\frac{2\Lambda}{\kappa} 
\,.
\ee

This geometry is also a solution of pure $R^2$ gravity, which is given by 
the action
\be
S_{f(R)} =\int d^4 x \, \sqrt{-g} \, f(R)
\ee
where \cite{Banijamali:2019gry} 
\be
f(R) = \frac{\kappa}{4\Lambda} \,  R^2 
\ee
and the effective scalar field is $\psi=f'(R)= \kappa R/( 2\Lambda)$. 
This theory does not have a Newtonian limit around a flat background 
\cite{PechlanerSexl66} (although it admits one around de Sitter 
backgrounds \cite{Nguyen:2023whv}), but approximates Starobinsky inflation 
in $f(R)=R+\alpha R^2$ gravity at high curvatures.

%%marker

Since 
\be
\nabla^c \psi=  \frac{2\psi_{\ast} {\delta_0}^c}{\kappa\tau_{\ast} \, r^2 
\Big( \frac{ \tau}{\tau_{\ast}} - 1 \Big)^3} 
\ee
and $\nabla^c \psi\nabla_c\psi<0$, the scalar field gradient is timelike but is future--oriented only for $\tau> \tau_{\ast}$, to which range we restrict.

The geometry was analyzed in \cite{Banijamali:2019gry}: the 
Ricci scalar and Ricci tensor squared are
\be
R = \frac {1}{\kappa \left(1-\tau/\tau_{\ast} \right)^2} \left( 2\Lambda 
\psi_{\ast}-\frac{4\omega}{\tau_{\ast}^2 r^2} \right) \,,
\ee
\begin{eqnarray}
R_{ab}R^{ab} &=& 
\frac {1}{\kappa \tau_{\ast}^4 r^4 \left(1-\tau/\tau_{\ast} \right)^4} 
\left( \frac{9}{\kappa \tau_{\ast}^2} -4 
+ \frac{ 8\Lambda r^2}{3} \right) \nonumber\\
&&\nonumber\\
&\, & +\frac{ \Lambda^2}{\kappa^2 \psi_{\ast}^2 } 
\left(1-\tau/\tau_{\ast} \right)^4 \,.
\end{eqnarray}
Both curvature scalars diverge as $r\to 0^{+}$ and as $\tau \to 
\tau_*^{+}$ which locate, respectively, timelike and spacelike  spacetime 
singularities. The former 
is again a naked central singularity in a background created by the 
cosmological constant, now morphed into the Jordan frame $V(\psi)$.  
Moreover, $G_\mathrm{eff}=1/\psi \to 0 $ as $\tau \to \tau_{\ast}^{+}$.

The effective temperature for this Brans--Dicke solution is given by
\be
{\cal KT} = \frac{1}{ 4\pi \sqrt{\kappa} \,  \tau_{\ast} r \left| 
1-\tau/\tau_{\ast} \right|} \,:
\ee
it diverges as $r\to0^{+}$ and as $\tau \to \tau_{\ast}^{+}$, while it 
vanishes asymptotically as $\tau\to +\infty$ and is monotonically 
decreasing in $\tau$--time. This behaviour is not as interesting as that 
of 
the Sultana solution~(\ref{Sultanametric2}), (\ref{Sultanaphi2}) and 
Eq.~(\ref{KTevolution}) describing the approach to the GR equilibrium 
misses the third term on its right--hand side since here $\Box\psi=0$.

\section{Conclusions}
\label{sec:5}
\setcounter{equation}{0}

In previous literature on the first--order thermodynamics of 
scalar--tensor gravity \cite{Faraoni:2018qdr, Faraoni:2021lfc, 
Faraoni:2021jri, Giusti:2021sku, Giusti:2022tgq, Faraoni:2023hwu, 
Miranda:2022wkz}, Eq.~(\ref{KTevolution}) was always studied in 
situations in which $\Box\psi=0$ to simplify the analysis. Two key ideas 
of this formalism emerged in this context: a)~spacetime singularities are 
``hot''; b)~the expansion of 3--space ``cools'' gravity 
\cite{Faraoni:2018qdr, Faraoni:2021lfc, Faraoni:2021jri, Giusti:2021sku, 
Giusti:2022tgq,Faraoni:2023hwu,Miranda:2022wkz}. Are these ideas valid 
only in the limited context $\Box\psi=0$ or are they generic?  Thus far, 
there is no known example in which the third term on the 
right--hand side of Eq.~(\ref{KTevolution}), proportional to 
$\Box\psi/\psi$, is non--vanishing and is 
allowed to play a role. It is interesting to learn how it can affect the 
evolution of $K{\cal T}$ and the Sultana solution~(\ref{Sultanametric2}), 
(\ref{Sultanaphi2}) for a conformally coupled scalar field in the Higgs 
potential~(\ref{Higgs}) allowed us to do just that. In this geometry, 
gravity is asymptotically Einstein gravity in the far past, then 
deviates from it but only by a finite extent ({\em i.e.}, by a finite 
${\cal KT}$), and returns to GR. This is a new behaviour due to the term 
$\frac{\Box\psi}{\kappa\psi}$: all example solutions previously 
examined show a monotonic relaxation to GR or 
departure from it, as exemplified by the Brans--Dicke geometry of  
Sec.~\ref{sec:4}.\footnote{It can be regarded as a Brans--Dicke cousin of 
the  Sultana solution since, like the latter, it is conformal to the 
Sultana--Wyman geometry.}$~$\\
$~$\hspace{0.345cm}In the general theory, it is not possible to predict {\em a priori} the sign of $\frac{\Box\psi}{\kappa\psi}$ and, therefore, its cooling or 
heating effect on gravity.  Specifying the functions $V(\psi)$ and 
$\omega(\psi)$, assuming $\nabla_c $ to be timelike, and 
restricting to $2\omega+3>0$ in order to avoid a 
phantom field $\psi$ seems to help, for then Eq.~(\ref{Boxpsi}) yields
\be 
\frac{\Box \psi}{\kappa\psi} = \frac{\psi V' - 2V +\omega' \left| 
\nabla^c\psi \nabla_c\psi \right| }{\kappa\psi \left| 2\omega+3\right| } 
\,, 
\ee 
where a prime denotes differentiation with respect to $\psi$ and, of 
course, $\psi>0$ to ensure that $G_\mathrm{eff}>0$ (we have inserted an 
absolute value to make explicit the sign of the term containing 
$\nabla^c\psi\nabla_c \psi<0$). For given $V(\phi)$ 
and $\omega(\phi)$ (often, $\omega$ is constant), it is possible to 
predict the sign of $\Box\psi/\psi$ but whether this term dominates or 
not, or whether it vanishes asymptotically cannot be decided {\em a 
priori}. To conclude, although the two key ideas of the first-order 
thermodynamics of scalar--tensor gravity are corroborated,  
the full range of possible behaviours of gravity is richer. 

\begin{acknowledgments}

This work is supported, in part, by the Natural Sciences \& Engineering 
Research Council of Canada (grant 2023--03234 to V.~F.) and by a Bishop's 
University Graduate Entrance Scholarship (N.~K.).

\end{acknowledgments}

\begin{appendices}

\section{Effective fluid components}
\label{Appendix:A}
\renewcommand{\theequation}{A.\arabic{equation}}
\setcounter{equation}{0}

In order to compute the various fluid quantities, it is useful to know 
that
\begin{eqnarray}
&&   \bigg( 1- \frac{4\,\pi\,\phi^2}{3} \bigg)^{-1} = \cosh^2(\alpha t) 
\,,\label{coshsquared}\\
&&\nonumber\\
&& \Box\phi=0 \,,\label{boxphi}\\
&&\nonumber\\
&& \Box \left(\phi^2 \right) = \frac{-2}{\kappa \,r^2\, \cosh^6 ( \alpha 
t)} \,,\label{boxphisquared}\\
&&\nonumber\\
&& \nabla^a\phi\nabla_a\phi =  \frac{-1}{ 
\kappa\,r^2\,\text{cosh}^6(\alpha t)}  \,,\label{nablaphi_nablaphi}\\
&&\nonumber\\
&& \nabla_a\nabla_b \big(\phi^2 \big) =  
{ \delta_a}^0 {\delta_b}^0 \, \frac{\left[ 2-6\sinh^2(\alpha 
t)\right] }{ \cosh^4(\alpha t) }  \nonumber\\ 
&&\nonumber\\
 &&\,\,\,\,\,\,\,\,\,\,\,\,\,\,\,\,\,\,\,\,\,\,\,\,\,\,\,\,\,\,\,\,\,\,\,
-\big({\delta_a}^0{\delta_b}^1 + {\delta_a}^1{\delta_b}^0\big)\, 
\frac{2\,\text{tanh}(\alpha t)}{\alpha\,r\,\cosh^2(\alpha t)} \, 
\nonumber\\
&&\nonumber\\
&&\,\,\,\,\,\,\,\,\,\,\,\,\,\,\,\,\,\,\,\,\,\,\,\,\,\,\,\,\,\,\,\,\,\,\,
-\big({\delta_a}^2{\delta_b}^2 + 
{\delta_a}^3{\delta_b}^3\,\sin^2\theta\big)\, 
\frac{2\,\tanh^2(\alpha t)}{\kappa\,\cosh^2(\alpha t)} \,, 
\nonumber\\
&&\label{nablanablaphisquared}\\
&&\nonumber\\
&& \nabla^a \phi \nabla^b \phi \nabla_a\nabla_b \phi = \frac{3\,\alpha 
\tanh(\alpha t)}{\kappa^2\,r^4\,\cosh^{10}( \alpha t) }  
\,\label{complexphicomponent}\\
&&\nonumber\\
&& \nabla^a \phi \nabla^b \phi \nabla_a\nabla_b \left( \phi^2 \right) = 
\frac{2 \left[ 1-3\sinh^2(\alpha t)  \right]}{ \kappa^2\, r^4\,\cosh^{12}( 
\alpha t) }\,.\label{complexphisquaredcomponent}
\end{eqnarray}

The energy density~(\ref{energydensity_definition}) for the 
Sultana metric~(\ref{Sultanametric}) and scalar field~(\ref{Sultanaphi}) is computed using Eqs.~(\ref{coshsquared}), (\ref{boxphisquared}), (\ref{nablaphi_nablaphi}) and (\ref{complexphisquaredcomponent}) in Eq.~~(\ref{energydensity_definition}), which yields
\begin{eqnarray}
&& \rho^{(\psi)} = \cosh^2(\alpha 
t)\Bigg\{\frac{\text{sech}^6(\alpha t)}{2\,\kappa 
\,r^2}+\frac{\Lambda\,\text{sech}^4(\alpha 
t)}{2\,\kappa} \nonumber\\ 
&&\nonumber\\
& & +\frac{1}{6}\bigg[ \bigg( \frac{2-6\,\text{sinh}^2(\alpha 
t)}{\kappa^2\,r^4\,\cosh^{12}(\alpha 
t)}\bigg)\bigg(\frac{-\kappa\,r^2}{\text{sech}^6(\alpha 
t)}\bigg)+\frac{2\,\text{sech}^6(\alpha t)}{\kappa\, r^2}\bigg]\Bigg\} 
\nonumber\\
&&\nonumber\\
&& = \frac{\cosh(2\alpha t)\, \text{sech}^4(\alpha t)}{2\, 
\kappa\, r^2} + \frac{\Lambda\, \text{sech}^2(\alpha t)}{2\, \kappa}
\end{eqnarray}
using $1+2\sinh^2 x=\cosh(2x)$.

The isotropic pressure~(\ref{isotropicpressure_definition}) of 
the Sultana solution~(\ref{Sultanametric}), (\ref{Sultanaphi}) is  
\begin{eqnarray}
&& P^{(\psi)} = \cosh^2(\alpha 
t)\Bigg\{\frac{\text{sech}^6(\alpha t)}{2\,\kappa\, 
r^2}-\frac{\Lambda\,\text{sech}^4(\alpha t)}{2\,\kappa}   \nonumber\\ 
&&\nonumber\\
& & +\frac{1}{18}\bigg[ \bigg(  \frac{2-6\,\sinh^2(\alpha 
t)}{\kappa^2\,r^4\,\cosh^{12}(\alpha t)}\bigg)\bigg(\frac{-\kappa\, 
r^2}{\text{sech}^6(\alpha t)}\bigg)-\frac{4\,\text{sech}^6(\alpha 
t)}{\kappa\, r^2}\bigg]\Bigg\}  \nonumber\\
&&\nonumber\\
&& =  \frac{\cosh(2\alpha t)\, \text{sech}^4(\alpha t)}{6\, \kappa\, 
r^2} - \frac{\Lambda\, \text{sech}^2(\alpha t)}{2\, \kappa}
 \,.\label{conformally_coupled_massive_scalar_field_pressure}
\end{eqnarray}
The viscous pressure for a conformally coupled scalar field can be found 
by subtracting the isotropic pressure of a minimally coupled scalar 
field, $P^{(\psi)}_{\xi = 0}$, from the isotropic pressure of a 
conformally coupled scalar field, $P^{(\psi)}_{\xi = 1/6}$, which is 
equivalent to the result of 
Eq.~(\ref{conformally_coupled_massive_scalar_field_pressure}):
\begin{eqnarray}
P_\mathrm{visc}^{(\psi)}  &=&  P^{(\psi)}_{\xi = 1/6} - 
P^{(\psi)}_{\xi = 0} = \frac{ \sinh^2(\alpha t) - 1}{3\,\kappa\, r^2 
\cosh^4(\alpha t)}  \,. 
\nonumber\\
&&
%&=& {\blue \bigg(\frac{1}{18-24\pi\phi^2}\bigg)\Bigg( 
%2\Box\big(\phi^2\big) +\frac{\nabla^a\phi\nabla^b\phi 
%\nabla_a\nabla_b\big(\phi^2\big)}{\nabla^e\phi\nabla_e 
%\phi}\Bigg)}\nonumber\\
%&&\nonumber\\
%&=& {\blue \frac{\cosh^2(\alpha 
%t)}{18}\Bigg[ \frac{-4\,\text{sech}^6(\alpha t)}{\kappa\, r^2}} 
%\nonumber\\
%&&\nonumber\\
%&+& {\blue \bigg(\frac{2-6\,\text{sinh}^2(\alpha 
%t)}{\kappa^2\,r^4\,\cosh^{12}(\alpha 
%t)}\bigg)\bigg(\frac{-\kappa\,r^2}{\text{sech}^6(\alpha t)}\bigg)\Bigg] } 
%\nonumber\\
%&&\nonumber\\
\end{eqnarray}

Moving to the heat flux for the conformally coupled scalar field, 
Eqs.~(\ref{heatflux_definition}), (\ref{coshsquared}), 
and~(\ref{nablaphi_nablaphi}) give 
\begin{eqnarray}
q_a^{(\psi)} &=& \frac{\cosh(\alpha 
t)}{6\,\sqrt{\kappa}\,r}\, 
\Big[-\nabla_a\nabla_0\big(\phi^2\big) 
+{\delta_a}^0\nabla_0\nabla_0\big(\phi^2\big)\Big] \,.\nonumber\\
&&\label{heatflux_simplify1}
\end{eqnarray}
The expression inside the square brackets of 
Eq.~(\ref{heatflux_simplify1}) is zero when  $a =0$, as it should be 
since $q_a^{(\psi)}$ is purely spatial. The second term 
inside these square brackets is zero for any other value of $a$, 
and only when $a=0,1$ is the first term non--zero. The use of 
Eq.~(\ref{nablanablaphisquared}) results in
\begin{eqnarray}
q_a^{(\psi)} &=& 
{\delta_a}^1 \, \frac{\cosh(\alpha  t)}{6\,\sqrt{\kappa}\,r} 
\, \Big[-\nabla_1\nabla_0\big(\phi^2\big)\Big]  = 
\frac{2{\delta_a}^1 \, \tanh(\alpha t)}{\sqrt{6}\, 
\kappa\,r^2\,\cosh(\alpha t)}  \,.\nonumber\\
&&\label{heatflux_simplify1}
\end{eqnarray}

%%%%%%%%%%%%%%%%%%%%%%

Since the anisotropic stress tensor $\pi_{ab}^{(\psi)}$ is purely spatial, 
we only compute its spatial components $\pi_{ij}$ ($i,j=1,2,3$). We have
\begin{widetext}
\begin{eqnarray}
\pi_{ij} &=& \frac{ \left( 1-\alpha^2\phi^2\right)}{-6\nabla^e 
\phi\nabla_e\phi} \left\{ -\frac{h_{ij}}{3} \, \nabla^e\phi \nabla_e\phi 
\left[ \Box \left( \phi^2 \right) -\frac{ \nabla^c \phi \nabla^d\phi 
\nabla_c \nabla_d \left( \phi^2 \right)}{\nabla^e\phi\nabla_e \phi} 
\right] \right.\nonumber\\
&&\nonumber\\
&\, & \left. +\left( \nabla^d\phi \nabla_d \phi \right) \nabla_i \nabla_j 
( \phi^2) - 
\nabla_j \phi \nabla^d \phi \nabla_i \nabla_d (\phi^2) -\nabla_i \phi 
\nabla^d \phi \nabla_j \nabla_d (\phi^2) 
 +\nabla_i \phi \nabla_j \phi \, \frac{ 
\nabla^c \phi \nabla^d\phi \nabla_c 
\nabla_d (\phi^2)}{\nabla^e\phi \nabla_e \phi} \right\} \,.
\end{eqnarray}
Using the fact that $\nabla_i\phi=0$ and computing 

\begin{eqnarray}
\Gamma^0_{ij} &=& \frac{\alpha \tanh(\alpha t)}{\kappa r^2} \left(\frac{2{\delta_i}^1\, {\delta_j}^1 }{1-2\Lambda r^2/3} + {\delta_i}^2 \,{\delta_j}^2 r^2  
+ {\delta_i}^3 \,{\delta_j}^3 \, r^2 \sin^2\vartheta \right) \,,\\
\nonumber\\
\nabla_i \nabla_j (\phi^2) &=& 2\phi \nabla_i \nabla_j \phi = -2\phi 
\, \Gamma^0_{ij} \dot{\phi} \nonumber\\
&&\nonumber\\
&=& -\frac{2 \tanh^2(\alpha t)}{\kappa r^2 \cosh^2(\alpha t)} \left(\frac{2{\delta_i}^1 \, {\delta_j}^1 }{1-2\Lambda r^2 /3} + {\delta_i}^2 \, {\delta_j}^2 r^2 + {\delta_i}^3 \, {\delta_j}^3 r^2 \sin^2\vartheta \right) \,,\\
\nonumber\\
\left(\nabla^d\phi \nabla_d \phi \right) \nabla_i \nabla_j (\phi^2) &=& \frac{-1}{\kappa r^2 \cosh^6(\alpha t) } \, 
\frac{-2\tanh^2(\alpha t)}{\kappa r^2 \cosh^2(\alpha t)} \left( \frac{2 {\delta_i}^1 \, {\delta_j}^1 }{1-2\Lambda r^2/3} + {\delta_i}^2 \, {\delta_j}^2 r^2 
+ {\delta_i}^3 \, {\delta_j}^3 r^2 \sin^2\vartheta \right) \nonumber\\
&&\nonumber\\
&=& \frac{2\tanh^2(\alpha t)}{\kappa^2 r^2 \cosh^{10}(\alpha t)} \, 
h_{ij} \,,\\
\nonumber\\
\nabla^d \phi  \nabla_i \nabla_d (\phi^2) &=& \frac{{\delta_i}^1}{\kappa 
r^3 \cosh^6(\alpha t)} 
\end{eqnarray}
and, putting everything together, we obtain
\begin{eqnarray}
\pi_{ij} &=& \frac{\kappa r^2 \cosh^6(\alpha t)}{6\left(1-\tanh^2(\alpha 
t) \right) } \left\{ 
\frac{h_{ij}}{3\kappa r^2 \cosh^6(\alpha t)} \left[ 
\frac{-2}{\kappa r^2 \cosh^2(\alpha t)} +\frac{2\left( 1-3\sinh^2(\alpha 
t) \right)}{\kappa r^2 \cosh^6(\alpha t)} \right] 
 +\frac{2\tanh^2(\alpha t)}{\kappa^2 r^4 \cosh^{10}(\alpha t)} 
\, h_{ij} \right\} \nonumber\\
&&\nonumber\\
&=& \frac{\kappa r^2 \cosh^8(\alpha t)}{6} \left[ \frac{ -2\tanh^2(\alpha 
t)}{\kappa^2 r^4 \cosh^{10}(\alpha t)} \, h_{ij} +\frac{2\tanh^2(\alpha 
t)}{\kappa^2 r^4 \cosh^{10}(\alpha t)} \right] =0 \,.
\end{eqnarray}
\end{widetext}

%%%%%%%%%%%%%%%%%%%%%%%

The shear tensor is \cite{Faraoni:2018qdr} 
\begin{widetext}
\begin{eqnarray}
\sigma_{ab} &=&  \left(-\nabla^e \phi \nabla_e 
\phi\right)^{-3 / 2}\Bigg[-\left(\nabla^e \phi  \nabla_e \phi\right) 
\nabla_a \nabla_b  \phi-\frac{1}{3}\left(\nabla_a \phi \nabla_b \phi-g_{a 
b} \nabla^c \phi \nabla_c \phi\right) \square \phi \nonumber\\
&&\nonumber\\
& \, & - \frac{1}{3}\left(g_{a b}+\frac{2 \nabla_a \phi \nabla_b 
\phi}{\nabla^e \phi \nabla_e \phi}\right) \nabla_c \nabla_d \phi \nabla^d 
\phi \nabla^c \phi+\left(\nabla_a \phi \nabla_c \nabla_b \phi+\nabla_b 
\phi \nabla_c \nabla_a \phi\right) \nabla^c \phi\Bigg] \,.
\end{eqnarray}
\end{widetext}
Since $\sigma_{ab}$ is purely spatial, one only needs to compute the 
spatial components, which straightforwardly gives $\sigma_{ab}=0$.  Using 
now  the four-velocity and Eq.~(\ref{nablaphi_nablaphi}), the expansion 
scalar~(\ref{expansion}) becomes 
\begin{eqnarray}
\Theta & = &  
\nabla_a\bigg(\frac{\nabla^a\phi}{\sqrt{-\nabla^b\phi\nabla_b\phi}}\bigg) 
\nonumber\\
&&\nonumber\\
&=&  \partial_a \bigg(\frac{{\delta_0}^a\,\text{sech}(\alpha t)}{ 
\sqrt{\kappa}\, r}\bigg) + 
\Gamma^a_{ab}\bigg(\frac{{\delta_0}^b\,\text{sech}(\alpha 
t)}{\sqrt{\kappa}\, r}\bigg) \nonumber\\
&&\nonumber\\
&=&  \frac{3\,\tanh (\alpha t)}{\sqrt{6}\,r\,\cosh(\alpha t)}  \,. 
\label{expansion_scalar_calculated} 
\end{eqnarray}

\end{appendices}

\end{document}